\documentstyle[12pt,epsfig]{article}
\def\beq{\begin{equation}}
\def\eeq{\end{equation}}
\def\bea{\begin{eqnarray}}
\def\eea{\end{eqnarray}}
\def\bq{\begin{quote}}
\def\eq{\end{quote}}

\def\bq{\begin{quote}}
\def\eq{\end{quote}}

\def\mpl{\ifmmode \overline M_{P}\else $\overline M_{P}$\fi}
\begin{document} 
\vspace*{-1in} 
\renewcommand{\thefootnote}{\fnsymbol{footnote}} 
\begin{flushright} 
TIFR/TH/01-10\\
hep-ph/0103055 \\ 
\end{flushright} 
\vskip 65pt 
\begin{center} 
{\Large \bf Constraining the Randall-Sundrum Model Using
         Diphoton Production at Hadron Colliders}\\
\vspace{8mm} 
{\large\bf 
         K.~Sridhar\footnote{sridhar@theory.tifr.res.in}
}\\ 
\vspace{10pt} 
{\sf Department of Theoretical Physics, \\
                     Tata Institute of Fundamental Research,\\  
                     Homi Bhabha Road, 
                     Bombay 400 005, India. } 

\normalsize
 
\vspace{20pt} 
{\bf ABSTRACT} 
\end{center} 
Virtual effects of gravitons in the production of diphotons at the 
upgraded Tevatron and at the LHC are analysed with the idea of probing 
the parameter space of the Randall-Sundrum (RS) model. It is shown that 
this process yields stringent constraints on the parameter space of the
RS model. We show that data on diphoton production from Tevatron Run II 
will be sensitive to a masses of the first graviton resonance in the range 
of 700-1150 GeV, while at LHC the mass range probed will be in the region 
of 3.5 -- 5.5 TeV.

\noindent

\vskip12pt 
\noindent 
\setcounter{footnote}{0} 
\renewcommand{\thefootnote}{\arabic{footnote}} 
 
\vfill 
\clearpage 

\setcounter{page}{1} 
\pagestyle{plain}


\noindent The physics of extra spacetime dimensions and its implications
for high energy physics and for astrophysics and cosmology has
attracted tremendous attention in recent years. The observable
universe is a dynamical hypersurface: a $D_3$-brane (or 3-brane) 
existing in a higher dimensional spacetime. The Standard Model (SM) fields
are localized on the brane but gravity can propagate in the 
bulk. In such scenarios, the gravity/string scale can be lowered down
from the Planck scale to the TeV scale \cite{string}. For high energy 
physics this is exciting because it provides fresh perspectives to the 
solution of the hierarchy problem and also suggests the discovery of new 
physics at TeV-scale colliders. 

The first realization of these ideas is the ADD scenario proposed by
Arkani-Hamed, Dimopoulos and Dvali \cite{dimo}, where, starting from
a higher dimensional theory, an effective four-dimensional theory
at a scale $M_S \sim {\rm TeV}$ is obtained. This is done by compactifying 
the extra dimensions to magnitudes which are large compared to the Planck 
length. The 
compactification radii in the ADD scenario can vary from a fermi to 
a millimetre, depending upon the number of large dimensions \cite{revadd}. 
While the ADD model does not run into any obvious conflict with
existing experimental data, it predicts large deviations from the SM in 
several phenomena which can be observed at existing and future
high-energy colliders \cite{phenoadd}. Thus, laboratory data can be used to
derive bounds on $M_S$ in the TeV range. Bounds
on this model have also been derived from cosmological and astrophysical
considerations and some these are considerably stronger than a TeV
\cite{astro}.

The main problem that one faces within the ADD model is the 
reappearance of disparate scales 
$viz.$, the string scale $M_S \sim 1$ TeV and the
compactification radius $R_c \sim \left( 10^{-16}~{\rm TeV} \right)^{-1}$.
The stability of these lrage dimensions is an undesirable feature
of this model and it was in an attempt to resolve this issue
that the Randall-Sundrum (RS) model originated \cite{rs}.
In its original form, the RS model is a five dimensional
model where the fifth dimension $\phi$ is compactified on a ${\bf
S}^1/{\bf Z}^2$ orbifold with a radius $R_c$ which is 
somewhat larger than the Planck length. At the orbifold fixed points, 
$\phi=0,\ \pi$, two 3-branes called the Planck brane and the TeV brane
are located. The SM fields are assumed to be localised on the TeV brane.
To get Poincar\' e invariance on the brane, it is necessary to fine-tune
the cosmological constants both on the brane and in the bulk. 
The model proposes a novel five-dimensional metric of the form
\begin{equation}
ds^2 = e^{-{\cal K}R_c\phi}\eta_{\mu\nu}dx^{\mu}dx^{\nu}~+~R_c^2d\phi^2 .
\end{equation} 
This metric is {\it non-factorizable} or $warped$ and the
exponential warp factor $e^{-{\cal K}R_c\phi}$ serves as a conformal factor 
for fields localised on the brane and this can be used to solve the 
hierarchy problem. The huge ratio
$\frac{M_P}{M_{EW}} \sim 10^{15}$ can be generated by the exponent
$\pi{\cal K}R_c$ which needs to be only of
{\cal O}(30). 
$R_c$ can be stabilised against quantum fluctuations
either by introducing an extra scalar field in
the bulk \cite{csaki, gold}, or by invoking supersymmetry \cite{bagger}. 

To derive the consequences of the RS model, a linearised gravity
approach is used where the curved metric is approximated by
fluctuations $h_{\mu\nu}$ about its Minkowski value. 
On compactification of the extra dimensions, a tower of
massive Kaluza-Klein (KK) excitations of the graviton, 
$h^{(\vec{n})}_{\mu\nu}$, result on the 3-brane. 
The interactions of these with the SM particles are given by:  
\begin{eqnarray} 
{\cal L}_{int} & = & -{1 \over \mpl} T^{\mu\nu}(x)
h^{(0)}_{\mu\nu}(x) -{e^{\pi {\cal K} R_c} \over \mpl} \sum_1^{\infty}
T^{\mu\nu}(x) h^{(n)}_{\mu\nu}(x) \ , 
\end{eqnarray} 
where
$\mpl=M_P/\sqrt{8\pi}$ is the reduced Planck mass and $T^{\mu\nu}$ is the
symmetric energy-momentum tensor for the observable particles on the
3-brane, computed using the flat space metric. The masses of the
$h^{(\vec{n})}_{\mu\nu}$ are given by 
\begin{eqnarray} 
M_{n} & = & x_n {\cal K} ~e^{-\pi {\cal K} R_c}
\end{eqnarray} 
where the $x_n$ are the zeros of the Bessel function $J_1(x)$ of order
unity \cite{gold}. The masses of the KK excitations are not evenly 
spaced in this model. The zero-mode in the tower of excitations
essentially decouples because of its weak coupling
but the couplings of the
massive RS gravitons are enhanced by the exponential $e^{\pi {\cal K}
R_c}$ leading to interactions of electroweak strength. 
The Feynman rules in this model are essentially the same as
those worked out\cite{grw, hlz} for the ADD case, except for the overall
warp factor in the RS case.

The basic parameters of the RS model are 
\begin{eqnarray}
m_0 & = & {\cal K} e^{-\pi {\cal K} R_c} \nonumber \\
c_0 & = & {\cal K}/M_P \label{m0c0}
\end{eqnarray}
where $m_0$ is a scale of the dimension of mass and sets the scale for 
the masses of the KK excitations, and $c_0$ is
an effective coupling. The interaction of massive KK gravitons with
matter can be written as 
\begin{equation} 
{\cal L}_{int}  = 
-\sqrt{8\pi} {c_0 \over m_0} \sum_n^{\infty} T^{\mu\nu}(x)
h^{(n)}_{\mu\nu}(x) \ . 
\end{equation} 
It is expected that the parameter $c_0$ lies in the range [0.01, 0.1]. This
is because the scale ${\cal K}$ is related to the curvature of the fifth 
dimension and so the upper bound on $c_0$ results if we want to avoid
strong curvature effects. But at the same time we would not want ${\cal K}$
to be too small as compared to $\mpl$, since that would introduce a new 
hierarchy. Values of
$m_0$ are determined in terms of ${\cal K} R_c \sim 10$, so that $m_0$
ranging from about a 100 GeV to a TeV are possible. We would like to
emphasise here that $m_0$ cannot become arbitrarily large in the RS model. 
This is because if $m_0$ becomes very large, it would require either 
${\cal K}$ to be large, or ${\cal K}R_c$ to be small (see Eq.~\ref{m0c0}). 
This results in a large curvature of the fifth dimension which makes it 
difficult to fine-tune the cosmological constants on the brane and the bulk to 
get a flat metric on the TeV brane. Consequently, the natural mass for the 
first graviton excitation is at most of the order of a few TeV.

It is interesting to ask what is the kind of collider phenomenology
that results with the RS model. Because of the fact that
the zero mode decouples, it is only the heavier modes one can
hope to detect in experiments. In the fortuitous circumstance that
these modes are within the reach of high-energy experiments, interesting
effects like resonance production can be observed, with the resonance
decaying within the detectors. If this is not the case and if the
the gravitons are heavier then the best strategy will be to look
for the virtual effects of the gravitons on observables measured
in high-energy collider experiments. Indeed, some of the phenomenology
of resonant production of the KK excitations and the virtual
effects have already been studied in processes like dilepton production 
at hadron colliders \cite{dhr}, $t \bar t$ production at hadron colliders 
\cite{lmrs} and in deep-inelastic scattering at HERA \cite{drs}. Novel 
effects like probing strong gravity via black-hole production at low energies 
have also been discussed in the context of the RS model \cite{giddings}.

In this letter, we study the virtual effects of the exchange of spin-2 KK 
modes, in the RS model, in diphoton production at the Tevatron and the LHC. 
In experiments, these photons are identified as isolated energy deposits
in the electromagnetic calorimeter. Selection cuts are then applied to
improve the signal-to-noise ratio. The D0 and CDF collaborations have
studied diphoton production in Run I of the Tevatron experiment.
The D0 collaboration has used these diphoton production data to 
derive constraints on the ADD model \cite{dipho}. The Run I studies
are however limited by statistics and for the purposes of the
present paper of limited interest. In this paper,
therefore, we concentrate on the production of diphotons at
Run II of the Tevatron and at the LHC. Because of the high statistics
that will be achieved in these experiments, it will be possible to perform
a detailed study of mass and angular distributions and 
derive possibly stringent constraints on the parameter space of the RS model.

The cross-sections for the $q\bar q \rightarrow \gamma \gamma$ and 
$gg \rightarrow \gamma \gamma$ subprocesses are \cite{dipho1,dipho2}:
\begin{eqnarray} 
{d\hat \sigma \over d\hat t}(q \bar q \rightarrow \gamma \gamma) 
& = & {2 \pi \alpha^2 Q_q^4 \over 3 \hat s^2} {1+{\rm cos}^2 \theta^*
\over 1-{\rm cos}^2 \theta^*}\\
&&  + {\alpha Q_q^2\over 96 \pi}{\rm Re}[C(x_s)] (1+{\rm cos}^2 \theta^*) 
+ {s^2 \over 24576 \pi }  |C(x_s)|^2 (1-{\rm cos}^4 \theta^*) ,
\end{eqnarray} 
and
\begin{eqnarray} 
{d\hat \sigma \over d\hat t}(gg \rightarrow \gamma \gamma) 
& = & 
+{s^2 \over 65536 \pi }  |C(x_s)|^2 (1+6{\rm cos}^2 \theta^*+{\rm cos}^4 
\theta^*) .
\end{eqnarray}
The SM box contribution $gg \rightarrow \gamma \gamma$ makes a very small
contribution at the Tevatron energy and is negligible \cite{dipho2}. At
the LHC energy, this box contribution is somewhat increased because
of the initial gluon flux but, as shown in Ref.~\cite{dipho2}, in spite
of this increase this contribution is an order of magnitude smaller
than the SM $q\bar q \rightarrow \gamma \gamma$ contribution for
diphoton invariant mass of 500 GeV and is more than two orders of
magnitude smaller for diphoton invariant mass greater than about
1750 GeV. On the other hand, the new physics effects dominate in
the large invariant mass bins and, therefore, in the invariant mass
region of interest the SM box contribution is negligible even for
the case of the LHC.

In the above equations, ${\rm cos} \theta^*$ is the scattering angle in
the partonic c.m. frame, $x_s \equiv \frac{\sqrt{\hat s}}{m_0}$ and
$C(x)$ is defined as
\begin{eqnarray}
C_(x) & = & \frac{32 \pi c_0^2}{m_0^4} \lambda(x)
\label{e1}
\end{eqnarray}
with
\begin{eqnarray}
\lambda(x_s)  & = & 
m_0^2 \sum_n \frac{1}{\hat s - M_n^2 + iM_n \Gamma_n} \ .
\end{eqnarray}
and the $M_n$ are the masses of the individual resonances and the
$\Gamma_n$ are the corresponding widths. 
The graviton
widths are obtained by calculating their decays into final states
involving SM particles. This gives
\begin{equation} 
\Gamma_n = m_0 c_0^2 x_n^3 \Delta_n
\end{equation} 
where 
\begin{equation}
\Delta_n = \Delta_n^{\gamma \gamma} + \Delta_n^{gg} 
         + \Delta_n^{WW} + \Delta_n^{ZZ} 
         + \sum_\nu \Delta_n^{\nu\nu} + \sum_l \Delta_n^{ll} 
         + \sum_q \Delta_n^{qq} 
         + \Delta_n^{HH} 
\end{equation}
and each $\Delta_n^{aa}$ is a numerical coefficient arising in the decay
$h^n \to a \bar a$. For the partial width $\Delta_n^{HH}$, we have
fixed $M_H = 250$ GeV in our numerical studies. 

Given the masses and the widths of the individual
graviton resonances, we have to sum over all the resonances to get the
value of $\lambda(x_s)$. We perform this sum numerically, using the
fact that the higher zeros of the Bessel function become evenly-spaced.
For a given value of $x_s = \frac{\sqrt{s}}{m_0}$, 
we retain all resonances which contribute with a
significance greater than one per mil, and treat the remaining KK modes as
virtual particles (in which case the sum can be done analytically).

Using the above sub-process cross-sections we can compute the diphoton
invariant mass distribution, $d\sigma/dM$ and the double differential
cross-section $d\sigma/dM d{\rm cos} \theta^*$ for $ \gamma \gamma$ 
production cross-section at the Tevatron and the LHC, by convoluting
with parton densities. For our numerical studies, we have used the
CTEQ4M parametrisations \cite{cteq} for the parton distributions. To
obtain the bounds on the $m_0$ -- $c_0$ parameter space, we compute
the cross-section for the diphoton production process in different
bins of $M$ (or $M$ and ${\rm cos}\theta^*$, for the case of the
double differential cross-section. Assuming Poisson-distributed data, 
we then compute the $\chi^2$ using the following expression:
\begin{equation} 
\chi^2 (m_0, c_0) = \sum_{i={\rm bins}}\biggl \lbrack 2 (n_i^{\rm th}
-n_i^{\rm obs}) + 2 n_i^{\rm obs}{\rm ln}({n_i^{\rm obs} \over n_i^{\rm th}})
\biggr \rbrack ,
\end{equation} 
where $n_i^{\rm obs}$ is taken to be the SM prediction for the number
of events in the bin $i$ and $n_i^{\rm th}$ is the prediction obtained
by adding the new physics contribution to the SM expectation. The $\chi^2$
so calculated is used to obtain a 95\% C.L. constraint on the $m_0$ -- $c_0$
parameter space. For Tevatron Run II, we use the following parameters:
$\sqrt{s}=2 TeV$, ${\cal L}= 2 {\rm fb}^{-1}$, bin size in $M$=80 GeV for 
$50<M<610$ GeV and $M=610 -- 1500$ GeV is combined into one bin, $y^{\gamma}<
\vert 1.2 \vert$ for each of the photons. When computing the double
differential cross-section $d\sigma/dM/d{\rm cos}\theta^*$, we use a bin
size of 0.2 for $-1 < {\rm cos}\theta^* < 1$.
For the LHC we use the following
parameters: $\sqrt{s}=14 TeV$, ${\cal L}= 100 {\rm fb}^{-1}$, bin size in 
$M$=200 GeV for $500<M<3250$ GeV and $M=3250 -- 5000$ GeV is combined into one 
bin, $y^{\gamma}< \vert 2.5 \vert$ for each of the photons. 
NLO QCD corrections to the SM contribution to the diphoton cross-section 
yields a K-factor of 1.3 (1.1) for Tevatron (LHC) \cite{owens}; we
assume same K-factor for the new physics contribution and simply
multiply the full cross-section by the K-factor to 
account for the effect of higher-order QCD corrections.

\begin{figure}[tbhp]
   \begin{minipage}{70mm}
\vspace*{0.3in}
  \begin{center}
    \leavevmode
    \epsfig{bbllx=171,bblly=256,bburx=469,bbury=606,
        file=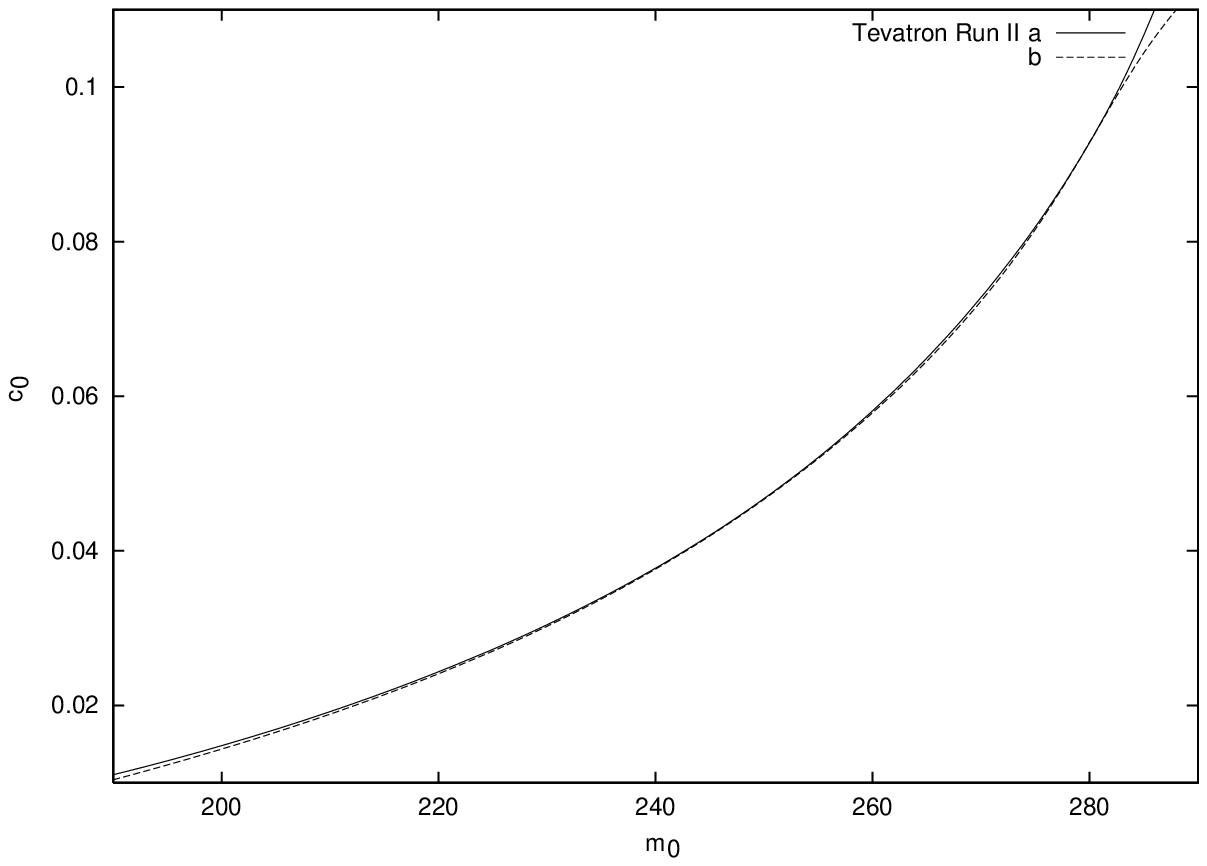,height=78mm,angle=270}
\vspace*{0.0in}
\hspace*{-2.0in}
\caption{\footnotesize\it Constraints on the $m_0-c_0$ plane
of the Randall-Sundrum model, using $\gamma \gamma$ production
at Run II of the Tevatron.}
\label{fig-pdk1}
   \end{center}
   \end{minipage}
\hfill
\begin{minipage}{70mm}
\vspace*{0.3in}
  \begin{center}
    \leavevmode
      \epsfig{bbllx=171,bblly=256,bburx=469,bbury=606,
          file=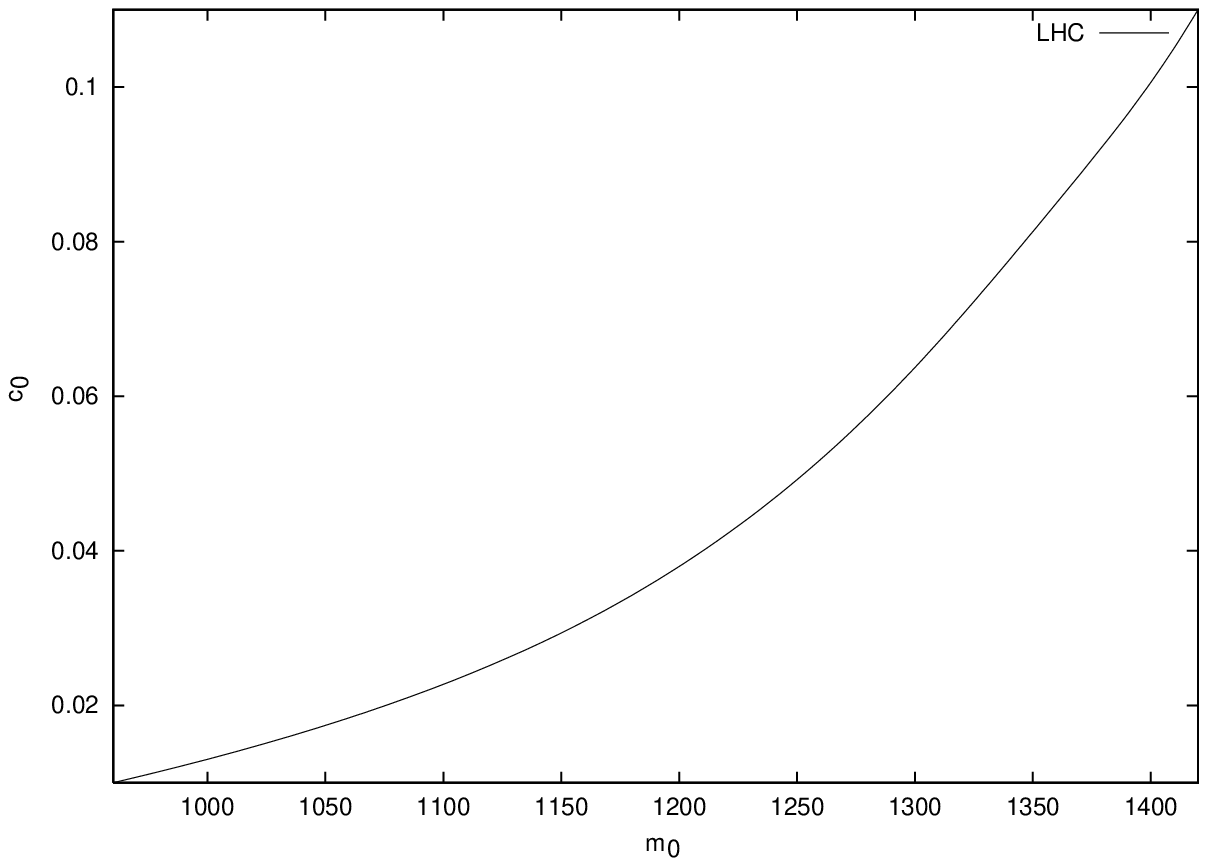,height=78mm,angle=270}
\vspace*{0.0in}
\hspace*{-2.0in}
\caption{\footnotesize\it Constraints on the $m_0-c_0$ plane
of the Randall-Sundrum model, using $\gamma \gamma$ production at the LHC.}
\label{fig-pdk2}
     \end{center}
   \end{minipage}
\end{figure}

We present the results of our computations in Fig.~1. In Fig~1($a$), we show 
the excluded region at 95\% C.L. in the $(m_0, c_0)$ plane obtained by 
computing the $\chi^2$ as described above. The two curves shown in the
figure are obtained by considering the $d\sigma/dM$ and the $d\sigma/dMd{\rm 
cos}\theta^*$ distributions. We find that the invariant mass distribution
already gives strong constraints but the inclusion of the information
on angular distribution does not improve these constraints. We find that
for values of $c_0$ between 0.01 and 0.1, the 95\% C.L. limits on $m_0$ that 
can be derived at the Run II of the Tevatron 
vary between 180 GeV and 290 GeV. This means that the first KK graviton 
resonance must lie above 700 GeV at the least. 
For the LHC, the constraints are shown in Fig.~1($b$). We find
that the accessible 95\% C.L. limits on $m_0$ for $0.01<c_0<0.1$ lie in the
range between 950 GeV and 1450 GeV, which means that the first graviton
resonance can be probed up to around 3.5~TeV. As discussed earlier,
the mass of the first graviton excitation is at most of the order
a few TeV, and it is an exciting prospect that this is the range of 
graviton masses is likely to be probed by the diphoton production process 
at the LHC.

To summarize, we have analysed the effects of the interactions of the
spin-2 Kaluza-Klein modes with SM fields in diphoton production at the
upgraded Tevatron (Run II) and at the LHC, in the context of the 
Randall-Sundrum model. By analysing the invariant mass distributions
of the photon pair, we derive 95\% C.L. limits on $m_0$ for $0.01<c_0<0.1$.
These limits are between 180 and 290 GeV for the Tevatron Run II and
between 950 and 1450 GeV for the LHC. The range of $m_0$ values 
that will be probed by this process at the LHC are such that there is
the exciting possibility of detecting the virtual effects at the
LHC or, conversely, non-observation
of these graviton modes at the LHC would seriously constrain the
Randall-Sundrum model, at least in its simplest form.

 

\begin{thebibliography}{999} 
 
\bibitem{string}
P.~Horava and E.~Witten, {\it Nucl. Phys.} {\bf B460} (1996) 506;
J.D.~Lykken {\it Phys. Rev.} {\bf D54} (1996) 3697;
E.~Witten, {\it Nucl. Phys.} {\bf B471} (1996) 135.

\bibitem{dimo} 
N.~Arkani-Hamed, S.~Dimopoulos and G.~Dvali, 
{\it Phys. Lett.} {\bf B249} (1998) 263;
I.~Antoniadis, N.~Arkani-Hamed, S.~Dimopoulos and G.~Dvali, 
{\it Phys. Lett.}  {\bf B436} (1998) 257. 

\bibitem{revadd} For a review of the ADD model, see 
I. Antoniadis and K. Benakli, {\it Int. J. Mod. Phys} {\bf A15} 
(2000) 4237; A. Perez-Lorenzana, Univ. of Maryland Preprint No.
UMD-PP-00-088 (2000) hep-ph/0008333.

\bibitem{phenoadd} For a review of the phenomenology of the ADD
model, see K. Sridhar, {\it Int. J. Mod. Phys} {\bf A15} (2000) 2397 
(hep-ph/0004053); S. Raychaudhuri, Talk given at the Sixth Workshop
on High-Energy Phenomenology, Chennai (India), January 2000. For
details of the phenomenology see Ref.~\cite{ours}.

\bibitem{astro} S. Cullen and M.~Perelstein, {\it Phys. Rev. Lett.} 
{\bf 83} (1999) 268; V.~Barger et al., {\it Phys. Lett.} {\bf B461} 
(1999) 34. 
 
\bibitem{rs} L. Randall and R. Sundrum, {\it Phys. Rev. Lett.} {\bf 83}
(1999) 3370.

\bibitem{csaki} C. Csaki, M. Graesser, L. Randall and J. Terning,
{\it Phys. Rev.} {\bf D 62} (1999) 045015, C. Csaki, M. Graesser and
G.D. Kribbs, Santa Cruz Preprint No. SCIPP-00-27, 
hep-th/0008151 (2000).

\bibitem{gold} W.D. Goldberger and M.B. Wise, Phys. Rev.  Lett. 
83 (1999) 4922; {\it Phys.Lett.} {\bf B475} (2000) 275.

\bibitem{bagger} R. Altendorfer, J. Bagger and D. Nemeschansky,
CITUSC-00-015, hep-th/0003117.

\bibitem{grw} 
G.~F.~ Giudice, R.~Rattazzi and J.~D.~Wells, 
{\it Nucl. Phys.} {\bf B544} (1999) 3. 

\bibitem{hlz} 
T.~Han, J.~D.~Lykken and R-J.~Zhang, {\it Phys. Rev.} {\bf D59} (1999) 105006.

\bibitem{dhr} H. Davoudiasl, J.L. Hewett and T.G. Rizzo, 
{\it Phys. Rev. Lett} {\bf 84} (2000) 2080; SLAC Preprint SLAC-PUB-8436 (2000) 
hep-ph/0006041.

\bibitem{lmrs} S. Lola, Prakash Mathews, Sreerup Raychaudhuri and K.~Sridhar, 
CERN Preprint No. CERN-TH-2000-275 (2000) hep-ph/0010010. 

\bibitem{drs} P. Das, S. Raychaudhuri and S. Sarkar, {\it JHEP} {\bf 0007:050}
(2000). 

\bibitem{giddings} S.B. Giddings and E. Katz, MIT Preprint No. MIT-CTP-3024
(2000) hep-th/0009176.

\bibitem{dipho} K. Cheung and G. Landsberg, 
{\it Phys. Rev.} {\bf D62} (2000) 076003; 
B.~Abbott et al., {\it Phys. Rev. Lett.} {\bf 86} (2001) 1156. 

\bibitem{dipho1} K. Cheung, {\it Phys. Rev.} {\bf D61} (2000) 015005.

\bibitem{dipho2} O.J.P. \' Eboli et al., {\it Phys. Rev.} {\bf D61} (2000) 
094007.

\bibitem{cteq} 
H.L.~Lai et al., {\it Phys. Rev.} {\bf D51} (1995) 4763 .

\bibitem{owens}
B.~Bailey, J.F. Owens and J. Ohnemus, {\it Phys. Rev.} {\bf D46} (1992) 2018.

\bibitem{ours}
Prakash Mathews, Sreerup Raychaudhuri and K.~Sridhar, {\it Phys. Lett.} 
{\bf B 450} (1999) 343; {\it Phys. Lett.} {\bf B 455} (1999) 115;
{\it JHEP} {\bf 0007:008} (2000). 

\end{thebibliography}
\end{document}